\documentclass[12pt]{article}

\usepackage{url}
\usepackage{bbm}
\usepackage{mathtools}
\usepackage{amssymb}
\usepackage{amsthm}
\usepackage{empheq}
\usepackage{latexsym}
\usepackage{enumitem}
\usepackage{eurosym}
\usepackage{dsfont}
\usepackage{appendix}
\usepackage{color} 
\usepackage[unicode]{hyperref}
\usepackage{frcursive}
\usepackage[utf8]{inputenc}
\usepackage[T1]{fontenc}
\usepackage{geometry}
\usepackage{multirow}
\usepackage{todonotes}
\usepackage{lmodern}
\usepackage{anyfontsize}
\usepackage{float}
\usepackage{pgfplots}
\usepackage[font=small,skip=0pt]{caption}
\usepackage{stmaryrd}

\pgfplotsset{compat=newest}

\definecolor{red}{rgb}{0.7,0.15,0.15}
\definecolor{green}{rgb}{0,0.5,0}
\definecolor{blue}{rgb}{0,0,0.7}
\hypersetup{colorlinks, linkcolor={red},citecolor={green}, urlcolor={blue}}
            
\makeatletter \@addtoreset{equation}{section}

\newtheorem{theorem}{Theorem}[section]

\newtheorem{lemma}[theorem]{Lemma}

\DeclareUnicodeCharacter{014D}{\=o}
\title{A note on Almgren-Chriss optimal execution problem with geometric Brownian motion\footnote{The ideas presented in this paper do not necessarily reflect the views or practices at Ritter Alpha LP. This work benefits from the financial support of the Chaires Analytics and Models for Regulation, Financial Risk and Finance and Sustainable Development. Bastien Baldacci gratefully acknowledge the financial support of the ERC Grant 679836 Staqamof. The authors would like to thank Robert Almgren, Peter Carr, Jim Gatheral, Dylan Possamai and Mathieu Rosenbaum for fruitful discussions. In particular, Peter Carr deserves warm thanks as this article would not exist without his inputs.}}
\author{\makebox[.9\textwidth]{Bastien {\sc Baldacci}}\\CMAP, Ecole Polytechnique\\ bastien.baldacci@polytechnique.edu \and Jerome  {\sc Benveniste}\\Ritter Alpha LP and NYU\\ ejb14@nyu.edu  }

\begin{document}

\maketitle
\begin{abstract}
We solve explicitly the Almgren-Chriss optimal liquidation problem where the stock price process follows a geometric Brownian motion. Our technique is to work in terms of cash and to use functional analysis tools. We show that this framework extends readily to the case of a stochastic drift for the price process and the liquidation of a portfolio.


\end{abstract}

\section{Introduction}
Optimal liquidation is a problem faced by a trader when he needs to liquidate a large number of shares. The trader faces a tradeoff between fast execution, reducing a risk related to price changes and slow execution, allowing to avoid high trading costs. Since the seminal paper by Almgren and Chriss \cite{almgren2001optimal}, various extensions of optimal liquidation problems have been studied, see for example~ \cite{alfonsi2010optimal,cartea2015optimal,gueant2012optimal}. The common framework to address this issue introduced in \cite{almgren2001optimal} assumes the following:
\begin{itemize}
    \item the efficient price process follows an arithmetic Brownian motion (ABM), 
    \item permanent market impact is linear,
    \item transaction costs are a linear function of the trading rate.
\end{itemize}
The execution of a large order is then formulated in discrete time as a tradeoff between expected costs and risk of the trading strategy, with variance as a risk measure. Under this framework, there exists a unique optimal liquidation strategy, which is a deterministic function of time and the initial position of the trader.  \\

Continuous versions of this problem have been considered, notably in \cite{forsyth2011hamilton,forsyth2012optimal}, where the author shows the ill-posedness of the mean-variance framework leading to time-inconsistent solutions. To overcome this issue, the authors suggest using alternative objective functions, particularly mean-quadratic variation. Under this choice, the authors solve a two-dimensional Hamilton-Jacobi-Bellman equation numerically. Moreover, in \cite{gueant2012optimal}, the authors consider the optimal execution problem with CARA utility objective function. To the best of our knowledge, there is no closed-form solution to the continuous version of the Almgren-Chriss framework with quadratic variation as a risk measure and geometric Brownian motion (GBM) assumption for the efficient price process. In \cite{gatheral2011optimal}, the authors solve a modified version of the problem with GBM, accounting for the risk with a linear function of the trading rate. \\ 

In this paper, we solve the optimal liquidation problem under the Almgren-Chriss framework in continuous time, in the case where the efficient price process follows a GBM, and the risk measure is quadratic variation. Motivated by \cite{collin2015prices}, we assume that trading costs are a quadratic function of the amount of cash, instead of shares. Thus, we reformulate the problem in terms of cash traded and derive in closed-form the optimal control of the trader liquidating his position. As the method is based on the resolution of a system of ODEs, it does not suffer from the curse of dimensionality. In particular, we show how to extend this framework to the case of the liquidation of a portfolio of $N$, possibly correlated, assets. It also enables us to treat the case where the return's drift is a stochastic process without any BSDE methods with a singular condition.  \\

The paper is organized as follows. In Section \ref{section_the_model}, we describe the Almgren Chriss framework in continuous time and reformulate the optimization problem in terms of cash. In Section \ref{section_solving_explicitly}, we obtain a closed-form solution of the Almgren Chriss framework with GBM for the efficient price process. Finally in Section \ref{section_numerical_results}, we present numerical applications under different market conditions. 

\section{The model}\label{section_the_model}

We define $(\Omega,\mathcal{F}_{t\in[0,T]},\mathbb{P})$ a filtered probability space, on which all stochastic process are defined, and a trading horizon is $T>0$.

\subsection{Almgren-Chriss framework in continuous time}
We rapidly recall the well-known Almgren-Chriss problem in continuous time. We consider the issue of the liquidation of $q_0 \in \mathbb{R}$ shares of a stock whose price at time $t$ is defined by $S_t$. The number of shares hold by the trader is defined by an absolutely continuous measurable process $q_t:=q_0 -\int_0^t \overset{.}{q}_s\mathrm{d}s$ where $(\overset{.}{q}_s)_{s\in[0,T]}$ is the trading rate, controlled by the trader. The transaction price is 
\begin{align*}
   \tilde{S}_t:=S_t + \frac{\lambda}{2} \overset{.}{q}_t + \gamma (q_t-q_0),
\end{align*}
where $\lambda,\gamma\in \mathbb{R}^+$ are constants related respectively to temporary and permanent price impact. Indeed, the term $\frac{\lambda}{2} \overset{.}{q}_t$ is the impact of trading $\overset{.}{q}_t$ shares at time $t$, whereas the term $\gamma (q_t-q_0)$ is the impact generated by the flow of transactions up to time $t$. In the original framework in discrete time, see \cite{almgren2001optimal}, and in most of the extensions in continuous time, see \cite{gueant2012optimal} for example, the price process follows an ABM. The number of shares hold by the trader satisfies the boundary condition $q_T=0$. Therefore, the cost of this strategy during the trading period is 
\begin{align*}
    \mathcal{C}(\overset{.}{q}):=\int_0^T \tilde{S}_t q_t\mathrm{d}t.
\end{align*}
Aiming at remedying the time inconsistency of the optimal strategies in the pre-commitment mean-variance framework, inspired by \cite{forsyth2012optimal}, we replace the variance by the quadratic variation in the penalty. The optimal execution problem consists in the optimization of a mean-quadratic variation objective function over the strategies $(\overset{.}{q}_t)_{t\in [0,T]}\in \mathcal{A}$ where
\begin{align*}
    \mathcal{A}:= \Big\{(\overset{.}{q}_t)_{t\in[0,T]}, \mathcal{F}_t-\text{measurable such that } \int_0^T \overset{.}{q}_s ds =q_0 \Big\}.
\end{align*}

The problem can be written as follows: 
\begin{align*}
 \sup_{v\in\mathcal{A}}\mathbb{E}\big[-\mathcal{C}(\overset{.}{q})-\frac{\kappa}{2}\langle \mathcal{C}\rangle_T \big],
\end{align*}
with $\kappa>0$ and
\begin{align*}
   \langle \mathcal C\rangle_T  := \int_0^T q_t^2 \mathrm{d}\langle S \rangle_t. 
\end{align*}
The use of quadratic variation leads to time-consistent strategies. Moreover, in contrast to the variance, quadratic variation takes into account the trajectory of liquidation. A direct integration by parts on $\mathcal C(q)$ gives 
\begin{align*}
    \mathcal{C}(q)=-q_0 S_0-\int_0^T q_t\mathrm{d}S_t + \frac{\lambda}{2} \int_0^T \overset{.}{q}_t^2\mathrm{d}t+\frac{\gamma}{2}q_0^2.
\end{align*}
Therefore, the problem writes as
\begin{align}\label{optimization problem AC classic 2}
 \sup_{\overset{.}{q}\in\mathcal{A}}\mathbb{E}\Big[\int_0^T q_t \mathrm{d}S_t - \frac{\lambda}{2}\int_0^T \overset{.}{q}_t^2\mathrm{d}t -\frac{\kappa}{2}\int_0^T q_t^2 \mathrm{d}\langle S \rangle_t \Big].
\end{align}

When the price process follows an ABM, Problem \eqref{optimization problem AC classic 2} boils down to a simple calculus of variations problem, which has been solved, for example, in \cite{gueant2012optimal}. The case where the dynamics are given by a GBM is more intricate. In \cite{gatheral2011optimal}, the authors consider it analytically intractable when a quadratic variation penalty is used. Moreover, in \cite{forsyth2012optimal}, the authors derive a numerical solution of \eqref{optimization problem AC classic 2} by solving the corresponding Hamilton-Jacobi-Bellman equation. Note that strategies under ABM assumption are good proxies of the ones under GBM assumption in period of low volatility. 

\subsection{Reformulation in terms of cash}

We now reformulate the optimal execution problem in terms of cash. We emphasize that we treat the very same problem as in \eqref{optimization problem AC classic 2}, except that we modify the transaction costs such that the penalty for $\overset{.}{q}_t$ becomes $\overset{.}{q}_t S_t$. \\

We assume that the price process follows a GBM:
\begin{align*}
    \mathrm{d}S_t=\sigma S_t\mathrm{d}W_t.
\end{align*}

Multiplying above and below by $S_t$, we obtain that
\begin{align*}
   \int_0^T q_t \mathrm{d}S_t = \int_0^T \theta_t \mathrm{d}y_t,
\end{align*}

where $\mathrm{d}y_t:=\sigma\mathrm{d}W_t$ is the return of the price process, and $\theta_t:= q_t S_t$ is the trader's position expressed in dollars. Moreover, the quadratic variation penalty has the form
\begin{align*}
    \frac{\kappa\sigma^2}{2}\int_0^T \theta_t^2 \mathrm{d}t.
\end{align*}

Applying Ito's formula, we derive that the cash position $\theta_t:=\theta_t^u$ has the following dynamics:\footnote{We write the superscript $u$ since $(u_t)_{t\in[0,T]}$ is the control process.}
\begin{align}\label{dynamic theta before change of variable}
  \mathrm{d}\theta_t^u = u_t \mathrm{d}t + \theta_t^u \mathrm{d}y_t = u_t \mathrm{d}t + \sigma \theta_t^u  dW_t, 
\end{align}
where $u_t= \overset{.}{q}_t S_t$ is the trading's rate in dollar at time $t$. \\

Recall that in the classical Almgren-Chriss framework \eqref{optimization problem AC classic 2}, trading costs are a quadratic function of the number of shares traded at time $t$ defined by $\overset{.}{x}_t$ (the second term in \eqref{optimization problem AC classic 2}). The only modification we make here is to assume that instantaneous costs are a quadratic function of the amount of cash. According to \cite{collin2015prices}, working with dollar holdings and returns is more consistent with common practice. We define the set of admissible control processes $(u_t)_{t\in[0,T]}$ as 
\begin{align*}
    \mathcal{A}:= \Big\{ (u_t)_{t\in[0,T]} \text{ measurable, s.t } \int_0^T | u_t | \mathrm{d}t < +\infty, \theta_T^u=0 \Big\}. 
\end{align*}
where the last condition ensures the complete liquidation of the trader's position at terminal time $T$. Following the problem formulation in terms of cash instead of shares, we consider the following mean-quadratic variation optimization problem:
\begin{align}\label{definition objective function}
    & \lim_{a\rightarrow+\infty} \sup_{u\in\mathcal{A}} \mathbb{E}\Big[\int_0^T -\Big(\frac{\lambda}{2}u_t^2+\frac{\kappa\sigma^2}{2}(\theta_t^u)^2\Big) \mathrm{d}t -\frac{a}{2}(\theta_T^u)^2 \Big].
\end{align}
The limit over $a>0$ aims at representing the singular condition $\theta_T^u=0$. Equation \eqref{definition objective function} can be seen as a classical linear-quadratic optimization problem, which is reduced to the resolution of a Riccati equation in dimension one. However such equations are not well suited for multidimensional extensions of this problem, that is to say the liquidation of a portfolio of $N$ assets. Furthermore when adding a possibly non-Markovian drift $(\alpha_t)_{t\in[0,T]}$ to the price process, one has to rely on BSDE methods to compute the optimal control.\\

Our method, developed in the next section, has several advantages. First, it enables us to solve the original Almgren-Chriss problem explicitly, under the GBM assumption, only by assuming that instantaneous costs are a function of the amount of cash. In addition to this, it applies to the case of a stochastic drift $(\alpha_t)_{t\in[0,T]}$ without using the BSDE framework. Finally, an explicit solution can be obtained in the case of the liquidation of a portfolio of $N$ possibly correlated assets.  \\

We solve in the next section Problem \eqref{definition objective function} under the dynamics \eqref{dynamic theta before change of variable} for the trader's position. We treat the non-zero drift case in Section \ref{stochastic drift section}.

\section{Solving explicitly the Almgren-Chriss problem with GBM}\label{section_solving_explicitly}

Throughout this section, we work on the following functional space:
\begin{align*}
\mathbb{H}^2 := \Big\{(v_t)_{t\in[0,T]}: \mathbb{E}\big[\int_0^T v_t^2\mathrm{d}t\big]<+\infty \Big\},
\end{align*}
with its associated inner product and norm 
\begin{align*}
\langle u,v \rangle_t = \mathbb{E}\Big[\int_0^t u_s v_s \mathrm{d}s\Big], \quad \| u \| =   \mathbb{E}\Big[\int_0^t u^2_s \mathrm{d}s\Big].
\end{align*}
We also define for all $t\in[0,T]$ the exponential martingale $M_t:=\exp\Big( \sigma W_t -\frac{\sigma^2}{2}t\Big)$ and the associated change of measure $\frac{\mathrm{d}\mathbb{Q}}{\mathrm{d}\mathbb{P}}\Big|_{\mathcal{F}_T}=M_T$. We begin with a lemma characterizing the trader's position.
\begin{lemma}\label{Lemma adjoint operator}
The unique solution of \eqref{dynamic theta before change of variable} is given by
\begin{align*}
   \int_0^t M_t M_s^{-1}u_s\mathrm{d}s. 
\end{align*}
For all $v\in \mathbb{H}^2$, we define the operator
\begin{align*}
   (Kv)_t := \int_0^t M_t M_s^{-1}v_s\mathrm{d}s.
\end{align*}
The adjoint process $(K^\star v)$ is equal for all $s\in[0,T]$ to
\begin{align*}
    (K^\star v)_s:= \int_s^T  \mathbb{E}^\mathbb{Q}[v_t|\mathcal{F}_s]\mathrm{d}t.
\end{align*}
\end{lemma}
The proof is given in Appendix \ref{proof Lemma adjoint operator} and relies on a straightforward application of Ito's formula. Therefore the optimization problem \eqref{definition objective function} can be rewritten, with a fixed $a>0$, as 
\begin{align}\label{Gateaux-differentiability}
   \sup_{u\in\mathcal{A}} -\frac{\lambda}{2}||u||^2-\frac{\kappa\sigma^2}{2}||Ku||^2 - \frac{a}{2}(Ku)_T^2. 
\end{align}
The problem is a supremum over a concave function of $u$, which is Gateaux-differentiable on $\mathbb{H}^2$. Thus first order condition gives:\footnote{See Appendix \ref{Section proof gateaux differentiability} for well-definedness of the first order condition.}
\begin{align}\label{first order condition}
\frac{\kappa\sigma^2}{\lambda}K^\star Ku + u + \frac{a}{\lambda}(Ku)_T = 0,
\end{align}
or equivalently
\begin{align} \label{equivalent first order condition}
    \frac{\kappa\sigma^2}{\lambda}\int_s^T \int_0^t \mathbb{E}^\mathbb{Q}[M_t M_\tau^{-1}u_\tau|\mathcal{F}_s]\mathrm{d}\tau\mathrm{d}t + u_s + \frac{a}{\lambda}M_T \int_0^T M_\tau^{-1}u_\tau \mathrm{d}\tau  = 0.
\end{align}
For all $(s,s_0)\in [0,T]^2$ such that $s\geq s_0$, we apply $\mathbb{E}^\mathbb{Q}[\cdot | \mathcal{F}_{s_0}]$ on both sides of \eqref{equivalent first order condition}. This leads to the following technical lemma.

\begin{lemma}\label{Main theorem 1}
We define $v(s):=\mathbb{E}^\mathbb{Q}[u_s|\mathcal{F}_{s_0}]$ such that $v(s_0)=u_{s_0}$, and assume that it is differentiable with respect to $s$.\footnote{It will be shown ex-post, by a direct verification argument, that $v(\cdot)$ is differentiable.} We also set 
\begin{align*}
& z(t) = e^{\sigma^2(t-s_0)}\theta_{s_0}+\int_{s_0}^t e^{\sigma^2(t-\tau)}v(\tau)\mathrm{d}\tau,
\end{align*}
where we recall that $\theta_{s_0}:=(Ku)_{s_0}$. \\

i) Equation \eqref{first order condition} can be rewritten 
\begin{align*}
    v(s)+ \frac{\kappa\sigma^2}{\lambda}\int_s^T z(t)\mathrm{d}t + \frac{a}{\lambda}z(T)  = 0.
\end{align*}

ii) The couple $(v,z)$ satisfies the following system of differential equations 
\begin{align*}
      \left\{
    \begin{array}{ll}
        v'(s)= \frac{\kappa\sigma^2}{\lambda}z(s) \\
        z'(s)=\sigma^2 z(s)+v(s), \\
    \end{array}
    \right.
\end{align*}
with boundary conditions 
\begin{align*}
      \left\{
    \begin{array}{ll}
        v(T)=-\frac{a}{\lambda}z(T) \\
        z(s_0)=\theta_{s_0}. \\
    \end{array}
    \right.
\end{align*}
\end{lemma}
Thus, the control problem \eqref{definition objective function} is reduced to the resolution of a linear system of ODEs with constant coefficients. We can now state our main theorem.

\begin{theorem}\label{Corollary explicit expression}
Consider the problem \eqref{definition objective function}, the optimal control is given explicitly for all time $t\in[0,T]$ by 
\begin{align*}
    u^\star_{t}= \theta_{t}^{u^\star}\Gamma(t), 
\end{align*}
where $\Gamma(\cdot)$ is a deterministic function of time defined in \eqref{Definition Gamma nu} and the optimal trader's position satisfies
\begin{align*}
    \theta_t^{u^\star} = \theta_0^{u^\star} \exp\big(\int_0^t (\Gamma(s) - \frac{\sigma^2}{2})\mathrm{d}s + \sigma W_t \big).
\end{align*}
\end{theorem}
The proof is included in the one of Theorem \ref{Corollary explicit expression stochastic}, where we prove a similar result in a more general framework by allowing a stochastic drift in the dynamics of the price process,  and is reported in Appendix \ref{proof corollary control}. The theorem shows that the optimal control is a linear function of the trader's position. Therefore, we find an \textit{aggressive in-the-money} selling strategy, similar to \cite{gatheral2011optimal}, in the sense that the trader liquidates faster when the stock price increases and conversely. This is illustrated in the following section. Moreover, the trader's position is a geometric Brownian motion, so that it always stays positive, in contrast to \cite{gatheral2011optimal}. As the function $\Gamma(t)\underset{t\rightarrow T}{\rightarrow} -\infty$ superlinearly, we have $\theta_t^u \underset{t\rightarrow T}{\rightarrow} 0$. 

\section{Numerical results}\label{section_numerical_results}

\begin{figure}[H]   
\begin{minipage}[c]{.46\linewidth}
     \begin{center}
             \includegraphics[width=\textwidth]{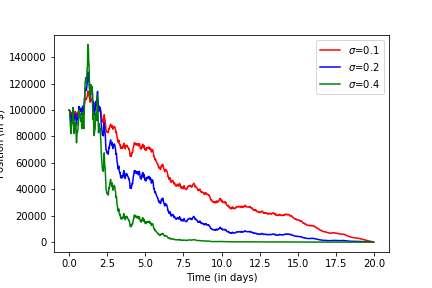}
             \vspace{-3mm}
             \caption{Evolution of the cash position with respect to time.}\label{dollar_position_sigma}
         \end{center}
   \end{minipage} \hfill
\begin{minipage}[c]{.46\linewidth}
     \begin{center}
             \includegraphics[width=\textwidth]{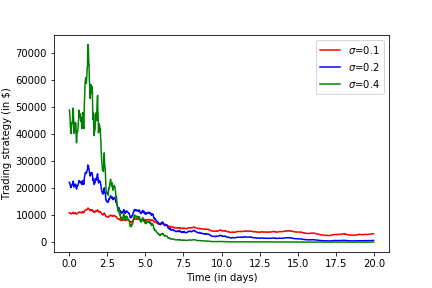}
             \caption{Trading strategy in cash with respect to time.}\label{dollar_strategy_sigma}
         \end{center}
   \end{minipage}   \hfill
\end{figure}

\begin{figure}[H]  
\begin{minipage}[c]{.46\linewidth}
     \begin{center}
             \includegraphics[width=\textwidth]{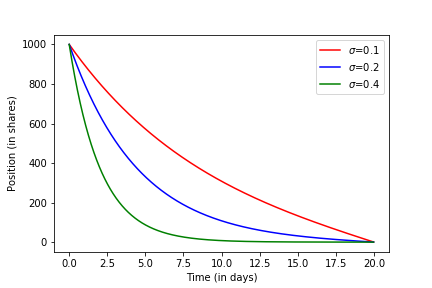}
             \caption{Evolution of the share's position with respect to time.}\label{shares_position_sigma}
         \end{center}
   \end{minipage}  \hfill
\begin{minipage}[c]{.46\linewidth}
     \begin{center}
             \includegraphics[width=\textwidth]{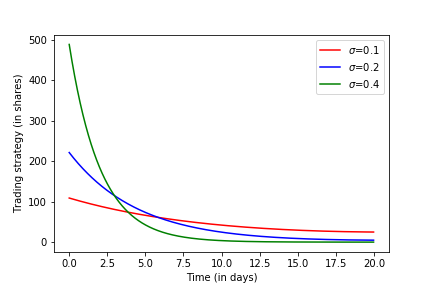}
             \caption{Trading strategy in shares with respect to time.}\label{shares_strategy_sigma}
         \end{center}
   \end{minipage}  \hfill
\end{figure}

\begin{figure}[H]  
\begin{minipage}[c]{.46\linewidth}
        \begin{center}
             \includegraphics[width=\textwidth]{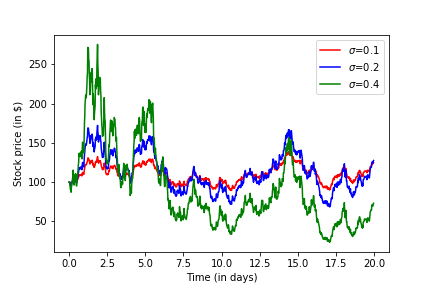}
             \caption{Evolution of the stock price with respect to time.}\label{stock price}
        \end{center}
    \end{minipage}    \hfill
\begin{minipage}[c]{.46\linewidth}
        \begin{center}
             \includegraphics[width=\textwidth]{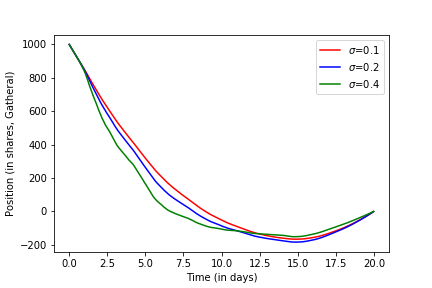}
             \caption{Evolution of the share's position with respect to time using \eqref{trading in shares gatheral}.}\label{shares_position_gatheral_sigma}
        \end{center}
   \end{minipage}
\end{figure}

We simulate one Brownian motion trajectory, and plot the corresponding stock price process, as well as trading strategy $(u_t^{\star})_{t\in[0,T]}$ and trader's cash position $(\theta_t^{\star})_{t\in[0,T]}$ and in shares $(\theta_t^\star/S_t)_{t\in[0,T]}$ for different values of $\sigma$. We take a stock with initial price $S_0=100\$ $ following a GBM without drift (whose trajectories for different values of $\sigma$ are in Figure \ref{stock price}), a portfolio of $10^3$ shares to liquidate over $T=20$ days, with $\lambda=\kappa=0.2$. In Figure \ref{dollar_position_sigma}, we see an increase of the cash position at the beginning, which can be misleading but is only due to the initial increase of the stock price process. This is also represented in the trading strategy of Figure \ref{dollar_strategy_sigma}, where we see that the trader liquidates his position faster when the stock process has a higher volatility. Figures \ref{shares_position_sigma} and \ref{shares_strategy_sigma} show the position and the trading strategy in terms of shares. We also compare in Figure \ref{shares_position_gatheral_sigma} our trading strategy in shares to the one in \cite{gatheral2011optimal}, which is defined as
\begin{align}\label{trading in shares gatheral}
    q_t^\star := \big(\frac{T-t}{T}\big)\Big(q_0-\frac{\kappa T}{4}\int_0^t S_u \mathrm{d}u\Big),
\end{align}
where $q_0=\frac{\theta_0}{S_0}$ is the initial number of shares hold by the trader, and $\mathrm{d}S_t=\sigma S_t\mathrm{d}W_t$. The trader still liquidates faster with a high volatility but his trading strategy, in this rather extreme regime, can go negative. 
\vspace{-10mm}
\vspace{0.00mm}

\begin{figure}[H]   
\begin{minipage}[c]{.46\linewidth}
     \begin{center}
             \includegraphics[width=0.97\textwidth]{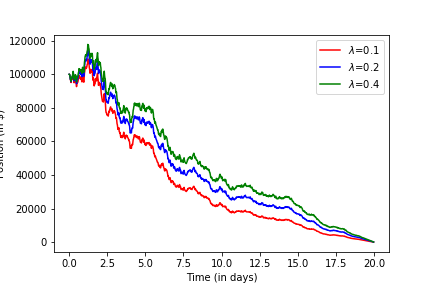}
             \caption{Evolution of the cash position with respect to time.}\label{dollar_position_lambda}
         \end{center}
   \end{minipage} \hfill
\begin{minipage}[c]{.46\linewidth}
     \begin{center}
             \includegraphics[width=0.97\textwidth]{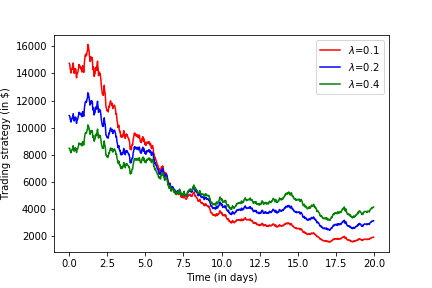}
             \caption{Trading strategy in cash with respect to time.}\label{dollar_strategy_lambda}
         \end{center}
   \end{minipage}   \hfill
\end{figure}

\begin{figure}[H]  
\begin{minipage}[c]{.46\linewidth}
     \begin{center}
             \includegraphics[width=0.97\textwidth]{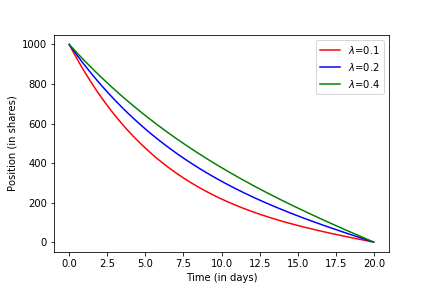}
             \caption{Evolution of the share's position with respect to time.}\label{shares_position_lambda}
         \end{center}
   \end{minipage}  \hfill
\begin{minipage}[c]{.46\linewidth}
     \begin{center}
             \includegraphics[width=0.97\textwidth]{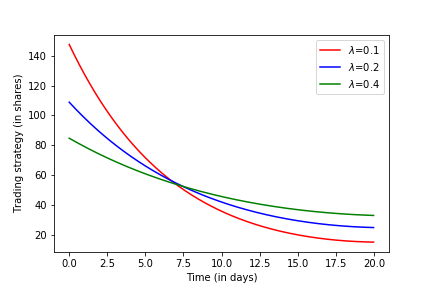}
             \caption{Trading strategy in shares with respect to time.}\label{shares_strategy_lambda}
         \end{center}
   \end{minipage}  
\end{figure}

We now fix $\sigma=0.1$ and $\kappa=0.2$. The various cases of the impact of the transaction costs $\lambda$ on the trader's behavior are represented in Figures \ref{dollar_position_lambda},\ref{dollar_strategy_lambda},\ref{shares_position_lambda} and \ref{shares_strategy_lambda}. Obviously, the price process is insensitive to a variation of $\lambda$. Moreover, the trading strategies in Figures \ref{dollar_strategy_lambda} and \ref{shares_strategy_lambda} are decreasing functions of $\lambda$ meaning that the trader liquidates his position using smaller sell orders when transactions costs are higher. This is also shown in the trader's position in Figures \ref{dollar_position_lambda} and \ref{shares_position_lambda}. 
\vspace{-10mm}

\begin{figure}[H]   
\begin{minipage}[c]{.46\linewidth}
     \begin{center}
             \includegraphics[width=0.97\textwidth]{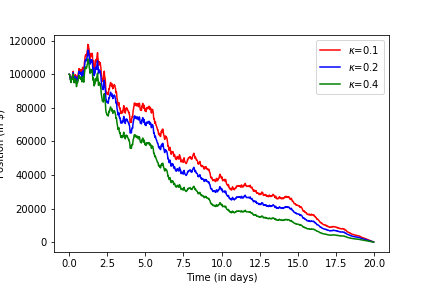}
             \caption{Evolution of the cash position with respect to time.}\label{dollar_position_kappa}
         \end{center}
   \end{minipage} \hfill
\begin{minipage}[c]{.46\linewidth}
     \begin{center}
             \includegraphics[width=0.97\textwidth]{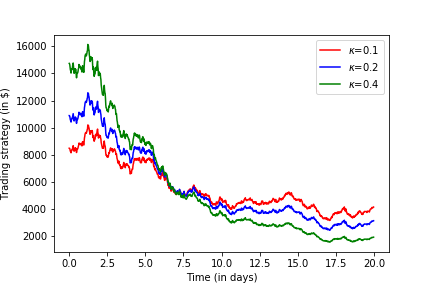}
             \caption{Trading strategy in cash with respect to time.}\label{dollar_strategy_kappa}
         \end{center}
   \end{minipage}   \hfill
\end{figure}

\begin{figure}[H]  
\begin{minipage}[c]{.46\linewidth}
     \begin{center}
             \includegraphics[width=0.97\textwidth]{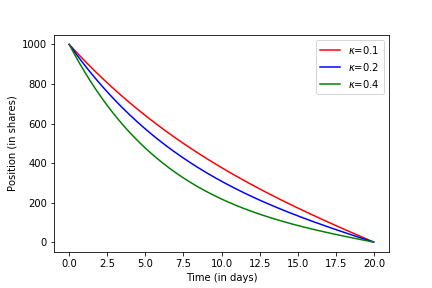}
             \caption{Evolution of the share's position with respect to time.}\label{shares_position_kappa}
         \end{center}
   \end{minipage}  \hfill
\begin{minipage}[c]{.46\linewidth}
     \begin{center}
             \includegraphics[width=0.97\textwidth]{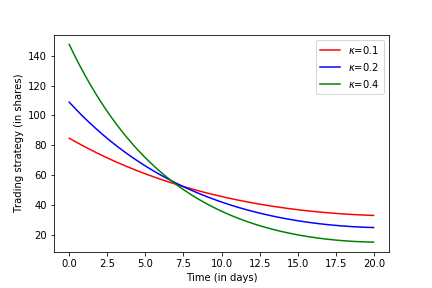}
             \caption{Trading strategy in shares with respect to time.}\label{shares_strategy_kappa}
         \end{center}
   \end{minipage}  
\end{figure}

Finally, we set $\sigma=0.1,\lambda=0.2$ and study the influence of the risk aversion parameter $\kappa$. In Figures \ref{dollar_strategy_kappa} and \ref{shares_strategy_kappa}, we see that a highly risk averse trader will liquidate faster than a low risk averse trader. This is shown in terms of his position in Figures \ref{dollar_position_kappa} and \ref{shares_position_kappa}.
\vspace{-10mm}

We now show how to extend our framework to the case of a stochastic drift for the price process and the liquidation of a portfolio of $N$ assets.






\section{Extensions of the model}\label{Section extensions}

\subsection{Stochastic drift}\label{stochastic drift section}

We now consider the case of a stochastic drift, that is we solve

\begin{align*}
    & \lim_{a\rightarrow+\infty} \sup_{u\in\mathcal{A}} \mathbb{E}\Big[\int_0^T \alpha_t \theta_t -\Big(\frac{\lambda}{2}u_t^2+\frac{\kappa\sigma^2}{2}(\theta_t^u)^2\Big) \mathrm{d}t -\frac{a}{2}(\theta_T^u)^2 \Big],
\end{align*}
where $(\alpha_t)_{t\in[0,T]}$ is a stochastic drift of the price process. We consider a slight modification of the problem where we neglect the part $\alpha_tS_t$ of the price's drift.\footnote{It can be shown that, if there exists $\eta>0$ such that $\sup_t |\alpha_t| < \eta$, then the trading strategy derived in this section is arbitrary closed (as a function of $\eta,T,\sigma$) to the optimal strategy without simplification of the drift.} Therefore, we simplify the dynamics of the price process, and assume
\begin{align}\label{dynamics stochastic drift simplified}
    \mathrm{d}\theta_t^u =u_t \mathrm{d}t + \sigma \theta_t^u \mathrm{d}W_t.
\end{align}
The first-order condition associated to this optimization problem writes as 
\begin{align*}
    \frac{\kappa\sigma^2}{\lambda}\int_s^T \int_0^t \mathbb{E}^\mathbb{Q}[M_t M_\tau^{-1}u_\tau|\mathcal{F}_s]\mathrm{d}\tau\mathrm{d}t + u_s + \frac{a}{\lambda}M_T \int_0^T M_\tau^{-1}u_\tau \mathrm{d}\tau  = \frac{1}{\lambda}\int_s^T \mathbb{E}^\mathbb{Q}[\alpha_t|\mathcal{F}_s]\mathrm{d}t.
\end{align*}

Then, the analogous of Equation \eqref{first order condition} can be rewritten 

\begin{align*}
    v(s)+ \frac{\kappa\sigma^2}{\lambda}\int_s^T z(t)\mathrm{d}t + \frac{a}{\lambda}z(T)  = \frac{1}{\lambda}\int_s^T \mathbb{E}^\mathbb{Q}[\alpha_t|\mathcal{F}_{s_0}]\mathrm{d}t.
\end{align*}
where the couple $(v,z)$ satisfies the following system of differential equations 
\begin{align*}
      \left\{
    \begin{array}{ll}
        v'(s)= \frac{\kappa\sigma^2}{\lambda}z(s)-\frac{1}{\lambda}\mathbb{E}^\mathbb{Q}[\alpha_s|\mathcal{F}_{s_0}]  \\
        z'(s)=\sigma^2 z(s)+v(s), \\
    \end{array}
    \right.
\end{align*}
with conditions 
\begin{align*}
      \left\{
    \begin{array}{ll}
        v(T)=-\frac{a}{\lambda}z(T) \\
        z(s_0)=\theta_{s_0}. \\
    \end{array}
    \right.
\end{align*}

We finally obtain the following theorem:

\begin{theorem}\label{Corollary explicit expression stochastic}
The optimal control at any time $t\in[0,T]$ is given by
\begin{align*}
    u^\star_{t}= \theta_{t}^{u^\star}\Gamma(t) + \nu(t), 
\end{align*}
where the optimal trader's position is defined as
\begin{align*}
    \theta_t^{u^\star}= H_t\int_0^t H_s^{-1}\nu(s)\mathrm{d}s
\end{align*}
with $\mathrm{d}H_t=\Gamma(t)H_t\mathrm{d}t+\sigma H_t\mathrm{d}W_t$, and $\Gamma(\cdot),\nu(\cdot)$ are deterministic functions defined in \eqref{Definition Gamma nu}. \\
\end{theorem}

The term $\nu(\cdot)$ is a linear function of both $\alpha_t$ and $\mathbb{E}^{\mathbb{Q}}[\alpha_T|\mathcal{F}_{t}]$, representing the influence of the drift on the optimal strategy. It is an increasing function of the drift $\alpha_t$ meaning that we aim at liquidating faster our position when the stock price increases. Moreover, it is a decreasing function of $\mathbb{E}^{\mathbb{Q}}[\alpha_T|\mathcal{F}_t]$: when the expected drift at the terminal time is high, the trader prefers to liquidate slower, waiting for a future stock price increase. As in the zero-drift case, we observe an \textit{aggressive in-the-money} selling strategy.

\subsection{Multi-dimensional case}

This model extends directly to the problem of optimal execution of a portfolio of $N$ assets. We define the return of the $i$-th asset as

\begin{align*}
    \mathrm{d}y_t^i=\sigma_i \mathrm{d}W_t^i,
\end{align*}
where $(W^1,\dots,W^N)$ are Brownian motions with non singular covariance matrix $\Sigma= (\sigma_i\sigma_j\rho^{i,j})_{1\leq i,j \leq N}$, $\sigma_i >0$ is the volatility of the $i$-th asset and $\rho^{i,j}$ is the correlation between the $i$-th and the $j$-th Brownian motion. The cash position of the trader with respect to the $i$-th asset is defined by
\begin{align}\label{Trader position i-th asset}
   \mathrm{d}\theta_t^{u,i}&=u_t^i\mathrm{d}t +\theta_t^{u,i}\mathrm{d}y_t^i =u_t^i \mathrm{d}t + \sigma_i \theta_t^{u,i}\mathrm{d}W_t^i,
\end{align}
where $(u_t^i)_{t\in[0,T]}$ is the trading rate on the $i$-th asset. Therefore the optimization problem \eqref{definition objective function} rewrites as
\begin{align*}
\lim_{a\rightarrow +\infty} \sup_{u\in\mathcal{A}} \mathbb{E}\bigg[\int_0^T \sum_{i=1}^N -\frac{\lambda}{2}(u_t^i)^2-\frac{\kappa}{2}\Big(\sum_{i=1}^N \sigma_i^2 (\theta_t^{u,i})^2\mathrm{d}t +\sum_{\substack{i,j=1 \\ i\neq j}}^N \rho^{i,j}\sigma_i\sigma_j \theta_t^{u,i}\theta_t^{u,j} \mathrm{d}t \Big) - \frac{a}{2}\sum_{i=1}^N (\theta_T^{u,i})^2 \bigg],
\end{align*}
where 

\begin{align*}
    \mathcal{A}:= \Big\{ (u_t^i)_{t\in[0,T], i\in\{1,\dots,N\}} \text{ measurable, s.t for all } i\in\{1,\dots,N\} \int_0^T | u_t^i | \mathrm{d}t < +\infty, \theta_T^{u,i}=0 \Big\}. 
\end{align*}
We define $(K^i u^i_t)_{t\in[0,T], i\in \{1,\dots,N\}}$ as the solution of the SDE \eqref{Trader position i-th asset}:
\begin{align*}
    (K^i u^i)_t = M_t^i \int_0^t (M_s^{i})^{-1}u^i_s \mathrm{d}s, 
\end{align*}
where $\frac{\mathrm{d}\mathbb{Q}^i}{\mathrm{d}\mathbb{P}}\Big|_{\mathcal{F}_T}=M_T^i:=\exp\big(\sigma_i W_T^i -\frac{(\sigma_i)^2 T}{2}\big)$. The adjoint operator is defined as
\begin{align*}
    (K^{\star i} u^i)_s = \int_s^T \mathbb{E}^{\mathbb{Q}^i} [u^i_t|\mathcal{F}_s] \mathrm{d}t.
\end{align*}
For a fixed $a>0$, the optimization problem rewrites  
\begin{align*}
    \sup_{u\in\mathcal{A}} -\frac{\lambda}{2}\| u \|^2 -\frac{\kappa}{2}\langle Ku, \Sigma Ku \rangle -\frac{a}{2}\sum_{i=1}^N (K^i u^i)^2_T.
\end{align*}
The first order condition gives the following system
\begin{align}\label{first order condition multidimensional asset}
    u^i + \frac{\kappa}{2\lambda}\big(2\sigma_i^2  K^{\star i}K^{i} u^i + \sum_{j\neq i}^N \rho^{i,j}\sigma_i\sigma_j K^{\star i}K^j u^j\big)+\frac{a}{\lambda}(K^i u^i)_T  =0, \quad i=1,\dots, N,
\end{align}
or equivalently for all $i=1,\dots,N$,
\begin{align*}
   u_s^i + \frac{\kappa}{2\lambda}\Big(2\sigma_i^2\int_s^T \int_0^t\big( \mathbb{E}^{\mathbb{Q}^i}[M_t^i (M_\tau^{i})^{-1}u_\tau^i|\mathcal{F}_s]\mathrm{d}\tau\mathrm{d}t + \sum_{j\neq i}^N \rho^{i,j}\sigma_i\sigma_j  \mathbb{E}^{\mathbb{Q}^i}[M_t^j (M_\tau^{j})^{-1}u_\tau^j|\mathcal{F}_s]\mathrm{d}\tau\mathrm{d}t\big)\Big)+\frac{a}{\lambda}(K^i u^i)_T=0. 
\end{align*}

For any $s\geq s_0$, apply $\mathbb{E}^{\mathbb{Q}^i}\big[\cdot|\mathcal{F}_{s_0}\big]$ on both sides of the equations. Simple but tedious computations lead to
\begin{align*}
   & \mathbb{E}^{\mathbb{Q}^i}\big[u^i_s|\mathcal{F}_{s_0}\big] + \frac{\kappa\sigma_i^2}{\lambda}\Big(\int_s^T e^{\sigma_i^2(t-s_0)}(K^iu^i)_{s_0}\mathrm{d}t + \int_s^T \int_{s_0}^t e^{\sigma_i^2(t-\tau)}\mathbb{E}^{\mathbb{Q}^i}\big[u_\tau^i|\mathcal{F}_{s_0}\big]\mathrm{d}\tau\mathrm{d}t\Big) \\
   & + \frac{\kappa}{\lambda}\sum_{j\neq i}^N \rho^{i,j}\sigma_i\sigma_j\Big( \int_s^T e^{\sigma_i\sigma_j\rho^{i,j}(t-s_0)} (K^ju^j)_{s_0} + \int_s^T \int_{s_0}^t e^{\sigma_i\sigma_j\rho^{i,j}(t-\tau)}\mathbb{E}^{\mathbb{Q}^j}\big[u_\tau^j|\mathcal{F}_{s_0}\big]\mathrm{d}\tau\mathrm{d}t\Big)\\
   & + \frac{a}{\lambda}\Big(e^{\sigma_i^2(T-s_0)}(K^iu^i)_{s_0} + \int_{s_0}^T e^{\sigma_i^2(T-\tau)}\mathbb{E}^{\mathbb{Q}^i}\big[u_\tau^i|\mathcal{F}_{s_0}\big]\mathrm{d}\tau\Big)  =0. 
\end{align*}

By denoting for all $i=1,\dots,N$, $v^i (s)=\mathbb{E}^{\mathbb{Q}^i}\big[u_s^i |\mathcal{F}_{s_0}\big]$, and $\theta_{s_0}^i=(K^i u^i)_{s_0}$ the system becomes
\begin{align*}
   & v^i(s) + \frac{\kappa\sigma_i^2}{\lambda}\Big(\int_s^T e^{\sigma_i^2(t-s_0)}\theta^i_{s_0}\mathrm{d}t + \int_s^T \int_{s_0}^t e^{\sigma_i^2(t-\tau)}v^i(\tau)\mathrm{d}\tau\mathrm{d}t\Big) \\
   & + \frac{\kappa}{\lambda}\sum_{j\neq i}^N \rho^{i,j}\sigma_i\sigma_j\Big( \int_s^T e^{\sigma_i\sigma_j\rho^{i,j}(t-s_0)} \theta^j_{s_0} + \int_s^T \int_{s_0}^t e^{\sigma_i\sigma_j\rho^{i,j}(t-\tau)}v^j(\tau)\mathrm{d}\tau\mathrm{d}t\Big) \\
   &  + \frac{a}{\lambda}\Big(e^{\sigma_i^2(T-s_0)}\theta^i_{s_0} + \int_{s_0}^T e^{\sigma_i^2(T-\tau)}v^i(\tau)\mathrm{d}\tau\Big) = 0. 
\end{align*}

We define 
\begin{align*}
& z^i(t):=e^{\sigma_i^2(t-s_0)}\theta^i_{s_0} + \int_{s_0}^te^{\sigma_i^2(t-\tau)}v^i(\tau)\mathrm{d}\tau, \\
& z^{i,j}(t):= e^{\sigma_i\sigma_j\rho^{i,j}(t-s_0)} \theta^j_{s_0} + \int_{s_0}^t e^{\sigma_i\sigma_j\rho^{i,j}(t-\tau)}v^j(\tau)\mathrm{d}\tau,
\end{align*}
and obtain for all $i=1,\dots,N$: 
\begin{align*}
   & v^i(s) + \frac{\kappa\sigma_i^2}{\lambda}\Big(\int_s^T z^i(t)\mathrm{d}t\Big) + \frac{\kappa}{\lambda}\sum_{j\neq i}^N \rho^{i,j}\sigma_i\sigma_j \int_s^T z^{i,j}(t)\mathrm{d}t + \frac{a}{\lambda}z^i(T)=0. 
\end{align*}

Therefore first-order condition \eqref{first order condition multidimensional asset} is equivalent to the system of differential equations
\begin{align*}
      \left\{
    \begin{array}{ll}
        v^{'i}(s)-\frac{\kappa\sigma_i^2}{\lambda}z^i(s)-\frac{\kappa}{\lambda}\sum_{j\neq i}^N \rho^{i,j}\sigma_i\sigma_j z^{i,j}(s)= 0   \\
        z^{'i}(s)=\sigma_i^2 z^i(s)+v^i(s) \\
        z^{'i,j}(s)=\sigma_i\sigma_j\rho^{i,j} z^{i,j}(s)+v^{j}(s), \\
    \end{array}
    \right.
\end{align*}

with initial conditions 
\begin{align*}
      \left\{
    \begin{array}{ll}
        v^{i}(T)= -\frac{a}{\lambda}z^i(T)   \\
        z^{i}(s_0)=\theta_{s_0}^i \\
        z^{i,j}(s_0)=\theta_{s_0}^j. \\
    \end{array}
    \right.
\end{align*}

We obtain a system of linear differential equations with constant coefficients. Thus, by noting that for all $i=1,\dots,N$ and $s_0\in[0,T]$, $v^i(s_0)=u_{s_0}^i$, we obtain the controls $u_t^i$ for all $t\in [0,T]$ and $i=1,\dots,N$ by solving this system of ODEs.

\section{Conclusion}
In this article, we present a way to solve the traditional Almgren-Chriss liquidation problem when the underlying asset is driven by a GBM. By working in terms of cash and using functional analysis tools, we can provide the optimal control of the problem explicitly. We provide an extension to the case of a GBM with stochastic drift and the liquidation of a portfolio of correlated assets. In particular, our method does not suffer from the curse of dimensionality.

\begingroup
\setcounter{section}{0}
\renewcommand\thesection{\Alph{section}}
\section{Appendix}

\subsection{Proof of Lemma \ref{Lemma adjoint operator}}\label{proof Lemma adjoint operator}

An application of Ito's formula gives 

\begin{align*}
    \mathrm{d}(Ku)_t= u_t \mathrm{d}t + \sigma (Ku)_t \mathrm{d}W_t,
\end{align*}
hence solving \eqref{dynamic theta before change of variable}. The adjoint of $K$ is the operator $K^\star$ such that for all $(u,v)\in\mathcal{A}$,
\begin{align*}
\langle Ku,v\rangle = \langle u, K^\star v\rangle.    
\end{align*}

Using Bayes formula, we have
\begin{align*}
   \langle Ku,v\rangle &= \mathbb{E}\Big[\int_0^T (Ku)_t v_t \mathrm{d}t \Big] \\
   &= \mathbb{E}\Big[\int_0^T \int_0^t M_t M_s^{-1}u_s v_t \mathrm{d}s \mathrm{d}t \Big] \\
   &= \mathbb{E}\Big[\int_0^T \int_s^T M_t M_s^{-1}u_s v_t \mathrm{d}t \mathrm{d}s \Big] \\
   &= \mathbb{E}\Big[\int_0^T u_s (K^\star v)_s \mathrm{d}s \Big] \\
   &= \langle u, K^\star v \rangle,
\end{align*}
where $(K^\star v)_s = \int_s^T M_t M^{-1}_s v_t\mathrm{d}t=\int_s^T \mathbb{E}^{\mathbb{Q}}[v_t|\mathcal{F}_s]\mathrm{d}t$. 

\subsection{Gateaux differentiability in \eqref{Gateaux-differentiability}}\label{Section proof gateaux differentiability}

We define the map $\Xi: \mathbb{H}^2 \rightarrow \mathbb{R}$ by
\begin{align*}
    \Xi(u)=-\frac{\lambda}{2}\| u \|^2 - \frac{\kappa\sigma^2}{2}\| Ku \|^2 - \frac{a}{2}(Ku)^2.
\end{align*}
As $K$ is a linear operator of $u\in \mathbb{H}^2$, and $\lambda,\kappa>0$, we deduce that $\Xi$ is continuous, strictly concave and Gateaux differentiable; with Gateaux derivative given, for any $h\in \mathbb{H}^2$, by
\begin{align*}
   \Xi(u)[h]= -\lambda \langle u, h \rangle -\kappa\sigma^2 \langle Ku, Kh \rangle - a (Kh) . 
\end{align*}
By setting $\Xi(u)[h]=0$, and using the definition of an adjoint operator,  we obtain Equation \eqref{first order condition}.

\subsection{Proof of Lemma \ref{Main theorem 1}}\label{proof main theorem}

By the fact that 
\begin{align*}
     \mathbb{E}^{\mathbb{Q}}\Big[\int_s^T \int_0^t \mathbb{E}^\mathbb{Q}[M_t M_\tau^{-1}u_\tau|\mathcal{F}_s]\mathrm{d}\tau\mathrm{d}t \big| \mathcal{F}_{s_0}\Big]& = \int_s^T \int_0^t \mathbb{E}^\mathbb{Q}[M_t M_\tau^{-1}u_\tau|\mathcal{F}_{s_0}]\mathrm{d}\tau\mathrm{d}t  \\
     & = \int_s^T \Big(\int_0^{s_0} M_\tau^{-1}u_\tau \mathbb{E}^{\mathbb{Q}}[M_t|\mathcal{F}_{s_0}] \mathrm{d}\tau + \int_{s_0}^t  \mathbb{E}^{\mathbb{Q}}[M_tM_\tau^{-1}u_\tau|\mathcal{F}_{s_0}] \mathrm{d}\tau\Big)\mathrm{d}t  \\
     & = \int_{s}^T\Big( e^{\sigma^2(t-s_0)}(Ku)_{s_0}\mathrm{d}t +  \int_{s_0}^t \mathbb{E}^{\mathbb{Q}}\big[\mathbb{E}^{\mathbb{Q}}[M_tM_\tau^{-1}u_\tau |\mathcal{F}_{s_0}]|\mathcal{F}_\tau\big]\mathrm{d}\tau \Big) \mathrm{d}t \\
     & =\int_{s}^T\Big( e^{\sigma^2(t-s_0)}(Ku)_{s_0}\mathrm{d}t +  \int_{s_0}^t e^{\sigma^2(t-\tau)}\mathbb{E}^{\mathbb{Q}}[u_\tau |\mathcal{F}_{s_0}]\mathrm{d}\tau \Big) \mathrm{d}t. 
\end{align*}
Condition \eqref{equivalent first order condition} can be rewritten 
\begin{align*}
    & \frac{\kappa\sigma^2}{\lambda}\int_{s}^T e^{\sigma^2(t-s_0)}(Ku)_{s_0}\mathrm{d}t + \frac{\kappa\sigma^2}{\lambda}\int_s^T \int_{s_0}^t e^{\sigma^2(t-\tau)}\mathbb{E}^{\mathbb{Q}}[u_\tau |\mathcal{F}_{s_0}]\mathrm{d}\tau \mathrm{d}t + \frac{a}{\lambda} e^{\sigma^2(T-s_0)}\theta_{s_0} \\
    & + \frac{a}{\lambda}\int_{s_0}^T e^{\sigma^2(T-\tau)}\mathbb{E}^{\mathbb{Q}}[u_\tau|\mathcal{F}_{s_0}] + \mathbb{E}^{\mathbb{Q}}[u_s|\mathcal{F}_{s_0}]=0,
\end{align*}

which proves the first statement of the theorem. We obtain the second point by a straightforward derivation of the functions $z$ and $v$. 

\subsection{Proof of Theorem \ref{Corollary explicit expression stochastic}}\label{proof corollary control}

The solution of the ODE for $z$ is given by 
\begin{align*}
    z(s)=\theta_{s_0}e^{\sigma^2(s-s_0)}+\int_{s_0}^s e^{\sigma^2(s-u)}v(u)\mathrm{d}u,
\end{align*}
and we can rewrite 
\begin{align*}
    v'(s)=\frac{\kappa\sigma^2}{\lambda}(\theta_{s_0}e^{\sigma^2(s-s_0)}+\int_{s_0}^s e^{\sigma^2(s-u)}v(u)\mathrm{d}u)-\frac{1}{\lambda}\mathbb{E}^{\mathbb{Q}}[\alpha_s|\mathcal{F}_{s_0}].
\end{align*}
Multiplying by $e^{-\sigma^2 s}$ on both sides we have
\begin{align*}
    e^{-\sigma^2s}v'(s)=\frac{k\sigma^2}{\lambda}(e^{-\sigma^2s_0}\theta_{s_0}+\int_{s_0}^s e^{-\sigma^2 u}v(u)\mathrm{d}u)-\frac{e^{-\sigma^2 s}}{\lambda}\mathbb{E}^{\mathbb{Q}}[\alpha_s|\mathcal{F}_{s_0}],
\end{align*}

and defining $w(s)=e^{-\sigma^2 s}v(s)$ we obtain
\begin{align*}
    w'(s)=\frac{k\sigma^2}{\lambda}(e^{-\sigma^2s_0}\theta_{s_0}+\int_{s_0}^s w(u)\mathrm{d}u)-\sigma^2 w(s)-\frac{e^{-\sigma^2 s}}{\lambda}\mathbb{E}^{\mathbb{Q}}[\alpha_s|\mathcal{F}_{s_0}].
\end{align*}

We note $y(s)=\int_{s_0}^s w(u)\mathrm{d}u$, satisfying the following differential equation
\begin{align*}
    y''(s)= \frac{\kappa\sigma^2}{\lambda}y(s)-\sigma^2 y'(s) + \frac{\kappa\sigma^2}{\lambda}e^{-\sigma^2s_0}\theta_{s_0}-\frac{e^{-\sigma^2 s}}{\lambda}\mathbb{E}^{\mathbb{Q}}[\alpha_s|\mathcal{F}_{s_0}].
\end{align*}

Solving this ODE without second member, we have 
\begin{align*}
    y(s)=C_1 e^{\gamma_1 s}+C_2 e^{\gamma_2 s},
\end{align*}
where $C_1,C_2\in\mathbb{R}$, $\gamma_1=\frac{-\sigma^2-\sqrt{\Delta}}{2},\gamma_2=\frac{-\sigma^2+\sqrt{\Delta}}{2}$, $\Delta=\sigma^2 (\sigma^2+4\frac{\kappa}{\lambda})>0$. A particular solution is given by  the function $y(s)=-\theta_{s_0} e^{-\sigma^2 s_0}+\frac{e^{-\sigma^2 s}}{\kappa\sigma^2}\mathbb{E}^{\mathbb{Q}}[\alpha_s|\mathcal{F}_{s_0}]$. The general solution is therefore given by:
\begin{align*}
    y(s)=C_1e^{\gamma_1 s}+C_2e^{\gamma_2 s}-\theta_{s_0}e^{-\sigma^2 s_0}+\frac{e^{-\sigma^2 s}}{\kappa\sigma^2}\mathbb{E}^{\mathbb{Q}}[\alpha_s|\mathcal{F}_{s_0}].
\end{align*}

To find $C_1,C_2$ we use the fact that $y(s_0)=0$ and $y'(T)=w(T)=e^{-\sigma^2 T}v(T)=-\frac{e^{-\sigma^2 T}a}{\lambda}z(T)$. Substituting the previous expression of $y$, and making $a\rightarrow +\infty$ to ensure liquidation at terminal time, we obtain 
\begin{align*}
    & C_1^{\infty}(s_0)=\beta^\infty(s_0)\bigg(e^{\gamma_2 T-\sigma^2 s_0}(\theta_{s_0}-\frac{\alpha_{s_0}}{\kappa\sigma^2})+e^{\gamma_2 s_0 -\sigma^2 T}\frac{\mathbb{E}^{\mathbb{Q}}[\alpha_T|\mathcal{F}_{s_0}]}{\kappa\sigma^2} \bigg), \\
    & C_2^{\infty}(s_0)=\beta^\infty(s_0)\bigg(-e^{\gamma_1 T-\sigma^2 s_0}(\theta_{s_0}-\frac{\alpha_{s_0}}{\kappa\sigma^2})-e^{\gamma_1 s_0 -\sigma^2 T}\frac{\mathbb{E}^{\mathbb{Q}}[\alpha_T|\mathcal{F}_{s_0}]}{\kappa\sigma^2} \bigg),
\end{align*}
where $\beta^\infty(s_0)=\frac{1}{e^{\gamma_1 s_0+\gamma_2 T}-e^{\gamma_1 T +\gamma_2 s_0}}>0$. Note that $v(s_0)=u_{s_0}=e^{\sigma^2 s_0}y'(s_0)$, which gives
\begin{align*}
    u_{s_0}= \theta_{s_0}\Gamma(s_0) + \nu(s_0),
\end{align*}
where
\begin{align}
 \begin{split}\label{Definition Gamma nu}
    & \Gamma(s_0):=\beta^{\infty}(s_0)(e^{\gamma_1 s_0 + \gamma_2 T}\gamma_1 - e^{\gamma_1 T +\gamma_2 s_0}\gamma_2),\\ 
    & \nu(s_0)=\beta^\infty(s_0)\Bigg(\gamma_1e^{(\gamma_1+\sigma^2)s_0}\Big(-\frac{\alpha_{s_0}}{\kappa\sigma^2}e^{\gamma_2 T - \sigma^2 s_0} + e^{\gamma_2 s_0 -\sigma^2 T} \frac{\mathbb{E}^{\mathbb{Q}}[\alpha_T|\mathcal{F}_{s_0}]}{\kappa\sigma^2}\Big) \\
    &\hspace{3em} + \gamma_2e^{(\gamma_2+\sigma^2)s_0}\Big(\frac{\alpha_{s_0}}{\kappa\sigma^2}e^{\gamma_1 T - \sigma^2 s_0} - e^{\gamma_1 s_0 -\sigma^2 T} \frac{\mathbb{E}^{\mathbb{Q}}[\alpha_T|\mathcal{F}_{s_0}]}{\kappa\sigma^2}\Big) \Bigg) - \frac{\alpha_{s_0}}{\kappa}.
\end{split}   
\end{align}

Substituting this expression in \eqref{dynamics stochastic drift simplified}, the trader's position becomes
\begin{align*}
    \mathrm{d}\theta_t^{u}= (\nu(t)+ \Gamma(t)\theta_t^u)\mathrm{d}t + \sigma \theta_t^u\mathrm{d}W_t.
\end{align*}

Therefore, we have the optimal position defined by 
\begin{align*}
    \theta_t^{u^\star}= H_t\int_0^t H_s^{-1}\nu(s)\mathrm{d}s,
\end{align*}
where $\mathrm{d}H_t=\Gamma(t)H_t\mathrm{d}t+\sigma H_t\mathrm{d}W_t$. The optimal control is finally given explicitly at any time $t\in[0,T]$ by 
\begin{align*}
    u^\star_{t}= \theta_{t}^{u^\star}\Gamma(t) + \nu(t). 
\end{align*}

\endgroup

\bibliographystyle{abbrv}
\bibliography{baldaccicarr.bib}

\end{document}